 \documentclass[final,5p,times,twocolumn]{elsarticle}

\usepackage{amssymb}
\usepackage{amsmath}
\newcommand{\Lagr}{\mathcal{L}}
\newcommand{\no}{\nonumber}
\newcommand{\ba}{\begin{eqnarray}}
\newcommand{\ea}{\end{eqnarray}}
\newcommand{\mev}{\textrm{ MeV}}

\usepackage[dvipsnames]{xcolor}

\begin{document}

\begin{frontmatter}



\title{Relevance of the coupled channels in the $\phi$p and $\rho^0$p Correlation Functions}

\author[inst1]{A. Feijoo}
\author[inst1]{M.  Korwieser}
\author[inst1]{L. Fabbietti}

\affiliation[inst1]{organization={Physik Department E62},
            addressline={Technische Universität München}, 
            city={85478 Garching},
            country={Germany}}


\begin{abstract}
The vector meson-baryon interaction in a coupled channel scheme is revisited within the correlation function framework. As illustrative cases to reveal the important role played by the coupled channels, we focus on the $\phi$p and $\rho^0$p systems given their complex dynamics and the presence of quasi-bound states or resonances in the vicinity of their thresholds. We show that the $\phi$p femtoscopic data provide novel information about a $N^*$ state present in the experimental region and anticipate the relevance of a future $\rho^0$p correlation function measurement in order to pin down the $S=0, Q=+1$ vector meson-baryon interaction as well as to disclose the characterizing features of the $N^*(1700)$ state.
\end{abstract}



\begin{keyword}
correlation function  \sep vector meson-baryon interaction \sep coupled channels \sep unitarized hidden gauge approach
\end{keyword}

\end{frontmatter}


\section{Introduction}
\label{sec:introduction}

With the advent of the recent $\phi$p correlation function (CF) analysis by the ALICE Collaboration \cite{ALICE:2021cpv}, an unprecedented access to the vector-baryon (VB) interaction at low energies arises in the $S=0, Q=+1$ sector. In particular, the ALICE collaboration extracted in \cite{ALICE:2021cpv} the $\phi$p scattering parameters (scattering length $a^{\phi p}_{0}$ and effective range $r^{\phi p}_{eff}$) employing the Lednický-Lyuboshits approach \cite{Lednicky:1981su}:
\ba
\label{scatt_ALICE}
a^{\phi p}_{0} &=&(0.85 \pm 0.48) + i\, (0.16 \pm 0.19) \,   \text{fm}  \\ \nonumber
r^{\phi p}_{eff} &=& 7.85 \pm 1.80  \, \text{fm}.
\ea
In light of the $a^{\phi p}_{0}$ imaginary part, compatible with $0$ within uncertainties, the authors naturally interpreted that the interaction was dominated by pure elastic $\phi$p interaction. With this assumption, in a subsequent study  \cite{Chizzali:2022pjd} of the $\phi$p CF \cite{ALICE:2021cpv}, a single channel phenomenological potential, fitted to $N-\phi$ Lattice data by the HAL QCD collaboration \cite{PhysRevD.106.074507}, was employed to explore the controversial existence of a $\phi-$p bound state. After solving the Schrödinger equation the binding energy obtained was within $[12.8,56.1]$ MeV which is below the lower edge of the  spectrum of binding energies, ranging from $1.00$ to $9.47$ MeV, provided by pertinent studies \cite{Huang:2005gw,Belyaev:2007yc,Sofianos_2010,Gao:2017hya,Sun:2022cxf}. Despite the previous evidence, the nature of this state remains ambiguous. A clear example of a competing interpretation comes from several studies based on extensions of Chiral Lagrangians to accommodate also vector mesons within a coupled channel (cc) unitary scheme \cite{Oset:2010tof,Khemchandani:2011et,Gamermann:2011mq,Garzon:2012np}. In all former approaches, this dynamically generated state, mostly consisting of a $K^*\Sigma$ molecule with sizeable couplings to $K^*\Lambda$ and $\phi$p channels, is located tens of MeV above the $\phi$p threshold.  \\

The study carried out in \cite{Oset:2010tof} represents one of the pioneering works incorporating the hidden gauge formalism \cite{Bando:1984ej,Bando:1987br,Meissner:1987ge,Harada:2003jx} into a cc unitary scheme. The corresponding VB scattering amplitudes in the sectors of $S=0,-1,-2$ exhibit rich structures caused by the presence of several states. However, given the approximations, such states appear as degenerate pairs of particles with spin parity $J^P=1/2^-, 3/2^-$. Some of the found states were related to known resonances while the others remain with an unclear connection to measured states due to the large associated dispersion from different experiments.

An important aspect of the chirally motivated formalisms is that the dominant leading-order term in the low energy approximation gives null $\phi$p and $\rho^0$p elastic direct transitions, which makes the dynamics of these processes rely on the cc effects (and /or on possible higher order corrections in the interaction). Such effects on the scattering amplitude are expected to be not negligible yet moderate,  a fact that leads one to consider the contributions from inelastic transitions as important ingredients in the $\phi$p and $\rho^0$p CFs. \\

The aim of the present study is twofold. On the one hand, the scattering amplitudes calculated following \cite{Oset:2010tof} and constrained exploiting the $\phi$p femtoscopic data are used to extract information about the state present close to the $\phi$p threshold and the scattering parameters. As a matter of fact, the constraining effect of the CF on the scattering amplitudes was already shown in \cite{Sarti:2023wlg} where the data extracted from the $K^-\Lambda$ pairs was used to determine the low energy constants of an effective Chiral Lagrangian expanded up to next-to-leading order. As a result, new insights on the molecular nature of the $\Xi(1620)$ and $\Xi(1690)$ states were obtained. On the other hand, we prove that the inelastic transitions cannot be underestimated or avoided in the analysis of certain CFs by means of delving into the relative weight of each scattering transition from any member of the cc basis to the measured vector-baryon. This point is clarified by explicitly discussing the contributions to the $\phi$p CF. Finally this study is extended to the $\rho^0$p pair extrapolating the scattering amplitudes to give a prediction for the CF as well as for the features of the $N^*(1700)$ state.\\

The manuscript is organized as follows. A brief description of the hidden gauge formalism within a cc unitary scheme and the computation of the CF is presented in Sec. \ref{sec:formalism}. This is followed by an explanation of the procedure to tune the parameters present in the approach in Sec. \ref{sec:results}, along with the discussion of the results for the $\phi$p and $\rho^0$p CFs, the pole content of the amplitudes, and the scattering parameters. Finally, the main conclusions are presented in Sec. \ref{sec:conclusions}.

\section{Formalism}
\label{sec:formalism}

\begin{table*}[h]
\begin{center}
\resizebox{10.4cm}{!} {
\begin{tabular}{|c|ccccccc|}
\hline
& & & & & & &\\[-1.5mm]
$C_{ij}$ &  $\rho^0 p$ &  $\rho^+ n$ &  $\omega p$ &  $\phi p$ & $K^{*+}\Lambda$ & $K^{*o}\Sigma^+$ & $K^{*+}\Sigma^0$  \\ [2mm] \hline
& & & & & & &\\[-1.5mm]
 $\rho^0 p$ & $0$ & $\sqrt{2}$ & $0$ & $0$ & $-\sqrt{3}/2$  & $1/\sqrt{2}$ & $-1/2$ \\ 
 $\rho^+ n$  &  & $1$ & $0$ & $0$ & $-\sqrt{3}/\sqrt{2}$   &  $0$ & $1/\sqrt{2}$ \\ 
 $\omega p$ &  &  &  $0$ & $0$ & $-\sqrt{3}/2$ & $-1/\sqrt{2}$ & $-1/2$ \\ 
 $\phi p$ &  &   &   & $0$ & $\sqrt{3}/\sqrt{2}$  & $1$ & $1/\sqrt{2}$ \\ 
$K^{*+}\Lambda$ &    &    &  &  & $0$ & $0$ & $0$ \\ 
 $K^{*o}\Sigma^+$&    &    &  &  & & $1$  & $\sqrt{2}$  \\ 
 $K^{*+}\Sigma^0$  &   &    &    &    &   &   & $0$ \\ [0mm] 
\hline
\end{tabular}
}
\caption{Values for the $C_{ij}$ coefficients of Eq.~(\ref{Kernel}). The coefficients are symmetric, $C_{ji} = C_{ij}$ .}
\label{Coeff_WT}
\end{center}
\end{table*}

In Refs. \cite{Sarkar:2010saz,Oset:2010tof}, it was demonstrated that the main contribution for the VB interaction comes from the computation of a t-channel diagram between the vector and the baryon with another vector meson acting as mediator. The hidden gauge formalism is followed to derive the interaction among the $3$ vector mesons present in one of the vertices, whose corresponding lagrangian reads as
\begin{eqnarray}
  \Lagr_{VVV} &=& ig \,\langle (V^\nu \partial_\mu V_\nu-\partial_\mu V^\nu V_\nu) V^\mu\rangle, 
 \label{Lag_3V} 
\end{eqnarray}
where the $SU(3)$-matrix $V_\mu$ incorporates the members of the $\rho$ nonet
\begin{equation}
\label{vmesfield}
V_\mu=\begin{pmatrix}
\frac{1}{\sqrt{2}}\rho^0+\frac{1}{\sqrt{2}}\omega & \rho^+ & K^{*+}\\
\rho^- & -\frac{1}{\sqrt{2}}\rho^0+\frac{1}{\sqrt{2}}\omega & K^{*0}\\
 K^{*-} & \bar{K}^{*0} & \phi\\
\end{pmatrix}_\mu,
\end{equation} 
and $g = M_V/2\,f$ ($M_V=800 \mev,~f=93 \mev$). The other vertex involving 2 baryons and the exchanged vector meson can be calculated by means of 
\begin{eqnarray}
  \Lagr_{BBV} &=& g \, (\langle \bar{B}\gamma_{\mu}[V^\mu , B]\rangle + \langle \bar{B}\gamma_{\mu}B \rangle \langle V^\mu\rangle ),
  \label{Lag_BBV} 
\end{eqnarray}
with the baryon octet fields entering into the equation through the $SU(3)$-matrix 
\begin{equation}
\label{Bfield}
B=\begin{pmatrix}
\frac{1}{\sqrt{2}}\Sigma^0+\frac{1}{\sqrt{6}}\Lambda & \Sigma^+ & p\\
\Sigma^- & -\frac{1}{\sqrt{2}}\Sigma^0+\frac{1}{\sqrt{6}}\Lambda & n\\
\Xi^- & \Xi^0 & -\frac{2}{\sqrt{6}}\Lambda\\
\end{pmatrix}  \ .
\end{equation} 
As it was already widely discussed in \cite{Oset:2010tof}, and the references therein, upon the approximation of neglecting the three-momentum of the external vector mesons compared to the characterizing $M_V$ mass of the vectors, the t-channel diagram is reduced to a contact interaction kernel which comes in terms of the product between the spatial components of the polarizations from the external vectors $\vec{\epsilon_i}\vec{\epsilon_j}$. This Weinberg-Tomozawa type kernel can be expressed in the relativistic form as 
\begin{equation}
\label{Kernel}
 V_{ij} = -\frac{1}{4f^2} C_{ij} \sqrt{\frac{M_i+E_i}{2M_i}}\sqrt{\frac{M_j+E_j}{2M_j}} (2\sqrt{s}{-}M_i{-}M_j) \, \vec{\epsilon_i}\vec{\epsilon_j}\,.
\end{equation}
In this work, the indices $(i,j)$ cover all possible channels in the $S=0, Q=+1$, namely $\rho^0$p, $\rho^+$n, $\omega$p, $\phi$p, $K^{*+}\Lambda$, $K^{*0}\Sigma^+$, and $K^{*+}\Sigma^0$. The mass and energy of the incoming (outgoing) baryons are denoted by $M_{i\,(j)}$ and $E_{i\,(j)}$, while $\sqrt{s}$ represents the total energy of the VB system in the center-of-mass (CM). The matrix of coefficients $C_{ij}$ can be found in Table~\ref{Coeff_WT}, from where it can be appreciated that the $\phi$p and $\rho^0$p direct elastic transitions are $0$.\\

Chiral unitary approaches have shown to be a powerful tool to treat the hadron scattering at energies close to resonances. These nonperturbative schemes prevent plain chiral perturbation theory from diverging and guarantee by construction the unitarity and analyticity of the scattering amplitude. In the present work, unitarity is implemented by solving the Bethe–Salpeter (BS) equation with coupled channels. Subsequently, with the fundamental input $V_{ij}$ calculated, the BS equation can be solved by factorizing the interaction kernel and the scattering amplitude out of the integral equation thereby leaving a simple system of algebraic equations to be solved. With each $V_{lm}$ vertex present in the different iterations of the BS loop diagrams, there is an implicit $\vec{\epsilon_l}\vec{\epsilon_m}$ factor involving the polarization of the vector mesons that also factorizes on-shell out of the loop integrals. Furthermore, since these states are tied to the polarization of the external vectors, one needs to sum over all polarizations of the virtual mesons ($\sum_{\text pol}\epsilon_k\epsilon_n=\delta_{kn}+q_kq_n/M_V^2$). Altogether, it leads to a correction in the propagator of $\vec{q}^2/3M_V^2$ that can be neglected consistent with the low energy approximation already assumed. Finally, the matrix form of the BS equation reads as 
\begin{equation}
T_{ij} ={(1-V_{il}G_l)}^{-1}V_{lj}\,,
 \label{T_algebraic}
\end{equation}
where $T_{ij}$ represents the scattering amplitude for a given starting $i$-channel and an outgoing $j$-channel, and $G_l$ is the loop function standing for a diagonal matrix with elements: 
\begin{equation} \label{Loop_integral}
G_l={\rm i}\int \frac{d^4q_l}{{(2\pi)}^4}\frac{2M_l}{{(P-q_l)}^2-M_l^2+{\rm i}\epsilon}\frac{1}{q_l^2-m_l^2+{\rm i}\epsilon} ,
\end{equation} 
where $M_l$ and $m_l$ are the baryon and vector masses of the transient $l$-channel. As this function diverges logarithmically, a dimensional regularization scheme is applied that gives the expression:
\ba
& G_l = &\frac{2M_l}{(4\pi)^2} \Bigg \lbrace a_l(\mu)+\ln\frac{M_l^2}{\mu^2}+\frac{m_l^2-M_l^2+s}{2s}\ln\frac{m_l^2}{M_l^2} + \no \\ 
 &     &\frac{q_{\rm cm}}{\sqrt{s}}\ln\left[\frac{(s+2\sqrt{s}q_{\rm cm})^2-(M_l^2-m_l^2)^2}{(s-2\sqrt{s}q_{\rm cm})^2-(M_l^2-m_l^2)^2}\right]\Bigg \rbrace ,  
 \label{dim_reg}    
\ea 
where the $q_{\rm cm}$ is the three-momentum modulus of the vector (baryon) in the CM framework. The subtraction constants (SCs) $a_l$ replace the divergence for a given dimensional regularization scale $\mu$, which is taken to be $630$~MeV in this work. Though the SCs are not determined, one can establish a natural size for them following the study presented in \cite{Oller:2000fj}, leading to a value around $-2$ for this study. To provide a certain versatility to the model, we will allow the SCs to vary within a plausible range. In principle, there are as many $a_l$'s as members of the coupled-channel basis, however, isospin symmetry arguments are frequently used to reduce the number of independent SCs. Hence, we consider $5$ such constants here: $a_{\rho N}$, $a_{\omega N}$,$a_{\phi N}$, $a_{K^{*}\Lambda}$ and $a_{K^{*}\Sigma}$. \\

The procedure described above is the standard one when the particles involved are stable, meaning that they have a negligible width. The presence of the $\rho$ and $K^*$ vectors with relatively large widths require a convolution of the loop function $G_l$ with the mass distribution whenever the iteration of the BS equation includes any of such mesons. This convolution mimics the effect of dressing the vector meson propagator inside the loop function. In practical terms, as in \cite{Oset:2010tof}, one has to replace the $G$-function in Eq.~(\ref{dim_reg}) by 
\ba
\Tilde{G}(s) &=& \frac{1}{N} \int_{(m_l-2\Gamma_l)^2}^{(m_l+2\Gamma_l)^2}dm^2 \Bigg( -\frac{1}{\pi} \Bigg ) Im \Bigg[ \frac{1}{m^2-m_l^2+im\Gamma(m)} \Bigg] \no \\ 
&\times& G_l(s,m^2,M_l^2) \, ,
\label{loop_conv} 
\ea
with the normalizing factor $N$ 
\ba
N=\int_{(m_l-2\Gamma_l)^2}^{(m_l+2\Gamma_l)^2}dm^2 \Bigg( -\frac{1}{\pi} \Bigg ) Im \Bigg[ \frac{1}{m^2-m_l^2+im\Gamma(m)} \Bigg]\, ,
\ea
and where the energy-dependent width is
\ba
\Gamma(m)=\Gamma_l \frac{m_l^2}{m^2}\Bigg( \frac{m^2-(m_{1}+m_{2})^2}{m_l^2-(m_{1}+m_{2})^2} \Bigg)^{3/2}\theta(m-(m_{1}+m_{2})) \, ,
\ea
which incorporates the decay widths $\Gamma_l$ ($\Gamma_\rho=149.77$ MeV, $\Gamma_{K^*}=48.3 $ MeV) and the corresponding decay products $m_1$ and $m_2$ for the given vector ($\rho \to m_1=m_2=m_\pi$ or $K^* \to m_1=m_K, m_2=m_\pi $).\\

The dynamically generated resonance states show up as pole singularities of the second Riemann sheet scattering amplitude at a complex value of $\sqrt{s}$ expressed as $z_R=M_R-{\rm i}\Gamma_R/2$, whose real and imaginary parts correspond to its mass ($M_R$) and the half width ($\Gamma_R/2$). The complex coupling strengths ($g_i$, $g_j$) of the resonance to the corresponding meson-baryon channels can be evaluated assuming a Breit-Wigner structure for the scattering amplitude in the proximity of the found pole on the real axis,
\begin{equation}\label{eq:pole}
T_{ij}(\sqrt{s})=\frac{g_i g_j} {(\sqrt{s}-z_R)} \,.
\end{equation}
In the present work, the use of mass distributions for the $\rho$ and $K^*$ mesons hampers a clear determination of the thresholds for the channels including such vectors thereby getting a fuzzy transition between the different Riemann sheets. In principle, this drawback becomes problematic when the resonance is located close to the nominal threshold, where the amplitude shape is distorted by the convolution and can even make the pole vanish. For these particular cases, as it was proved in \cite{Oset:2010tof}, one can obtain, with very good approximation, the location of the poles by examining the amplitudes obtained without the mass distribution. Then, in order to calculate the couplings, one can take the amplitude properly convoluted with the mass distribution and employ Eq.~(\ref{eq:pole}) as follows
\begin{equation}\label{eq:pole2}
|g_i|^2=\frac{\Gamma_R}{2} \sqrt{|T_{ii}|^2} \,.
\end{equation}
This equation allows, up to a global phase, the determination of the $g_i$ of the channel that most strongly couples to the resonance. Here, $|T_{ii}|$ is the position of the amplitude maximum in the real axis at $\sqrt{s}=M_R$. The other couplings can be derived similarly with  
\begin{equation}\label{eq:pole3}
g_j=g_j\frac{T_{ij}(\sqrt{s}=M_R)} {T_{ii}(\sqrt{s}=M_R)} \,. \\
\end{equation}

Finally, the two-particle CF in multi-channel systems for a measured channel $i$ is given by the modified Koonin-Pratt formula  \cite{Lisa:2005dd,Haidenbauer:2018jvl,Vidana:2023olz}
\begin{equation}
\label{CF}
C_i(k^*) = \sum_{j} \omega_{j} ^{\rm prod.} \int d^3 r^* S_{j} (r^*) |\psi_{ji} (k^*,r^*)|^2 \, ,
\end{equation}
which is preceded by the summation over all possible produced $j$-pairs connected with the final $i$-channel by theory ($j= \rho^0 p,~ \rho^+ n,~ \omega p,~ \phi p,~ K^{*+}\Lambda,~ K^{*0}\Sigma^+,~ K^{*+}\Sigma^0$). The variables $k^*$ and $r^*$ represent the relative momentum and distance between the two particles observed in the pair rest frame, respectively. The contributions from the transitions are scaled by the production weights $\omega_j^{\rm prod.}$ that were evaluated with the same data-driven method employed in~\cite{ALICE:2022yyh}, which was measured in the same high-multiplicity dataset as the one considered here. Since this study is exploratory we only quote the found central values without a full uncertainty analysis of the production weights, from~\cite{ALICE:2022yyh} the precision is expected to be $O(10\%)$. The corresponding values needed to compute the $\phi$p and $\rho^0$p CFs are displayed in Table~\ref{production_weights}. The probability of emitting a $j$-pair at a relative distance $r^*$ is taken as a single Gaussian with source size $R=1.08$~fm in accordance to \cite{ALICE:2021cpv},
\begin{equation}
S_j(r^*)=\frac{1}{\sqrt{4\pi}R_j^3}exp\Bigg( -\frac{{r^*}^2}{4R_j^2} \Bigg) \, .
\end{equation}
The last element to compute the CF is the relative wave function 
\ba
\label{psi}
&&\psi_{ji} (k^*,r^*)= \, \delta_{ij} j_0(k^*r^*) \nonumber \\ 
&&+\int_{q \le q_{\rm cut}} d^3 q j_0(qr^*)G_j(\sqrt{s},q)T_{ji}(\sqrt{s},k^*,q) \, ,
\ea
which describes the transition from a given channel $j$ to the final channel $i$ and incorporates the corresponding scattering amplitude in the common element to all transitions, while the first term, only affecting the elastic transition, is the zero-order Bessel function that stands for the asymptotic wave function. Note that the scattering amplitude $T_{ji}$ can be factorized on-shell out of integral and, then, one can proceed with the integration as in \cite{Vidana:2023olz,Molina:2023jov}, where the cutoff $q_{\rm cut}=850$ MeV has been established by calculating the average value of the equivalent cutoff to the corresponding loop calculations via dimensional regularization (method described in Ref~\cite{Oller:2000fj}).

The last step to establish a fair comparison with the genuine $\phi-$p CF, $C^{\text gen}_{\phi p}(k^*)$, obtained in the experimental analysis \cite{ALICE:2021cpv} is to multiply the theoretical CF that comes out from Eq.~(\ref{CF}) by a global normalizing factor $N_D$, which has to be extracted from the fit to femtoscopic data, in the following way 
\begin{equation}
\label{CF_gen}
C^{\text gen}_{\phi p}(k^*) = N_D C_{\phi p}(k^*).
\end{equation}
\begin{table}[h]
\begin{center}
\begin{tabular}{c|cc}
\hline
channel$-j$ & $\omega_j^{prod}$ ($\rho^0 p$ CF) & $\omega_j^{prod}$  ($\phi p$ CF)   \\ [2mm] \hline
$\rho^0 p$ & $1$ & $6.24$  \\ 
 $\rho^+ n$  & $0.95$ & $5.94$  \\ 
$\omega p$ & $0.92$ & $5.77$ \\ 
 $\phi p$ & $0.16$ &  $1$  \\ 
 $K^{*+}\Lambda$ & $0.10$  & $0.65$  \\ 
$K^{*o}\Sigma^+$ & $0.067$  &  $0.41$  \\ 
 $K^{*+}\Sigma^0$  & $0.069$  &  $0.43$  \\ [0mm] 
\hline
\end{tabular}
\caption{Values of the production weights for $\phi$p and $\rho^0$p CFs.}
\label{production_weights}
\end{center}
\end{table}

\section{Procedure and Results}
\label{sec:results}

As already mentioned, we use the $\phi$p genuine CF experimental data to determine the parameters present in the approach to extract information about the $\phi$p scattering parameters and perform an analysis of the pole content we obtain from the constrained scattering matrix. With this model, one can give predictions for the case of $\rho^0$p. In what follows, we discuss the different models developed.
In principle, we showed that the potential models should depend on $6$ parameters namely:
\begin{itemize}
  \item $5$ SCs ($a_{\rho N}$, $a_{\omega N}$,$a_{\phi N}$, $a_{K^{*}\Lambda}$ and $a_{K^{*}\Sigma}$) that have to take values around their natural size $-2$, it seems physically reasonable to constrain their values within the range $[-4,-1]$.
  \item $N_D$ normalizing factor, according to experimental standards this value should be around $1$, therefore, we allow this parameter to be between $0.8$ and $1.2$.
\end{itemize}
However, to check the validity of our approach, we decided to use a first model, called Pure theoretical, where all the parameters are taken as fixed values according to the theoretical estimations that can be seen in the first column of Table~\ref{tab:outputs_fits}. Next, we consider a second model, called Bootstrap, whose parameters and their uncertainties are obtained (as well as of every other quantity, e.g. scattering lengths) by means of the bootstrap technique \cite{Efron:1986hys} exploiting the 13 data points available in the $\phi$p femto region ($0<k^*<500$ MeV/c) of \cite{ALICE:2021cpv}. The uncertainty bands associated to the CFs are estimated taking the maximum and minimum CF values at each considered $k^*$ when varying the parameters $1\sigma$ from the uncertainties obtained with the bootstrap method. A fully fledged evaluation of the uncertainties, going beyond the estimate given here, related to the employed model must take into account the experimental precision of the source size. This is planned for a follow-up study once the measured $\phi$p and $\rho^0$p CFs are available. 
\begin{table}[h]
\centering
\begin{tabular}{lcc}
\hline \\[-2.5mm]
     & \textbf{Pure theoretical}  &  \textbf{Bootstrap}      \\                             
\hline \\[-2.5mm]
$a _{\rho N}$    & $-2$  \, (fixed)   &  $-2$  (fixed)  \\
$a_{\omega N}$   &  $-2$ \, (fixed)   & $ -3.04 \pm 0.73$       \\
$a_{\phi N}$     &  $-2$ \, (fixed)   &  $ -3.15 \pm 0.37$       \\
$a_{K^{*}\Lambda}$ &  $-2$ \, (fixed) &  $ -1.98 \pm 0.08$      \\
$a_{K^{*}\Sigma}$   &  $-2$ \, (fixed) &  $ -1.95 \pm 0.08$      \\
$N_{D}$           & $1$  \, (fixed)   &  $ 0.988\pm 0.004 $  \\
\hline
\end{tabular}
  \caption{Values of the parameters for the different models (see details in the manuscript).} 
\label{tab:outputs_fits}  
\end{table}
\begin{table}[h]
  \caption{Covariance matrix for the fitting parameters.} 
  \label{tab:cov_matrix}
\centering
\begin{tabular}{cccccc|r}
$a _{\rho N}$ & $a_{\omega N}$ & $a_{\phi N}$ & $a_{K^{*}\Lambda}$ & $a_{K^{*}\Sigma}$ & $N_D$ & $r_{ij}$\\  
[2mm] \hline
 $1.000$ & $0.000$ & $0.000$ & $0.001$ & $0.000$ & $0.000$ & $a _{\rho N}$\\
& $1.000$ & $-0.683$ & $-0.081$ & $-0.392$ & $-0.360$ & $a_{\omega N}$ \\
& &$1.000$ & $-0.056$ & $-0.283$ & $-0.267$ & $a_{\phi N}$  \\
& & &$1.000$ &$-0.361$ & $0.327$ & $a_{K^{*}\Lambda}$ \\
& & & &$1.000$ & $0.555$ & $a_{K^{*}\Sigma}$ \\
& & & & &$1.000$ & $N_D$ \\
\label{cov_matrix}
\end{tabular}
\end{table}

Before proceeding with the bootstrap analysis, we decided to check the covariance matrix, displayed in Table~\ref{tab:cov_matrix}, to investigate and understand possible correlations between the parameters of the model. Once the preliminary fits were performed and before inspecting the covariance matrix, we spotted sizeable errors, compared to the allowed range, for the $a _{\rho N}$, $a_{\omega N}$ and $a_{\phi N}$:
\ba
\delta (a _{\rho N})\approx \pm 2.5 \, ; \, \, \, \delta (a_{\omega N}), \delta (a_{\phi N})\approx \pm 2 \
\ea
Besides, when looking at the first row of Table~\ref{tab:cov_matrix}, it can be appreciated that the $a _{\rho N}$ SC is completely uncorrelated to the other parameters of the model. This fact, together with the large associated uncertainty, points out the impossibility of constraining this parameter from this data set. Consequently, we decided to fix its value to the theoretical natural size $a _{\rho N}=-2$. Regarding the other two problematic parameters, by checking the correlation coefficient $r_{23}$ in Table~\ref{tab:cov_matrix}, it can be observed that these two SCs are strongly correlated, thus finding a plausible explanation for the large uncertainties associated with them. Therefore, they were kept as free parameters taking values within the allowed interval. Eventually, these considerations leave the model depending on only 5 parameters.

In conclusion, we present the average values of the parameters and their uncertainties obtained with the bootstrap method in the right column of Table~\ref{tab:outputs_fits}. As significant outputs, we would like to stress the reduction of the uncertainties in the previous $a_{\omega N}$ and $a_{\phi N}$, as well as the small error we obtained for $a_{K^{*}\Sigma}$ since it plays a fundamental role in generating the $N^*$ state close to $\phi$p threshold discussed below. The small uncertainty for $a_{K^{*}\Lambda}$ comes as a consequence of the strong correlation it has to the $a_{K^{*}\Sigma}$ SC (see $r_{56}$ in Table~\ref{tab:cov_matrix}).
\begin{figure}[!ht]
\centering
\includegraphics[width=3.8 in ]{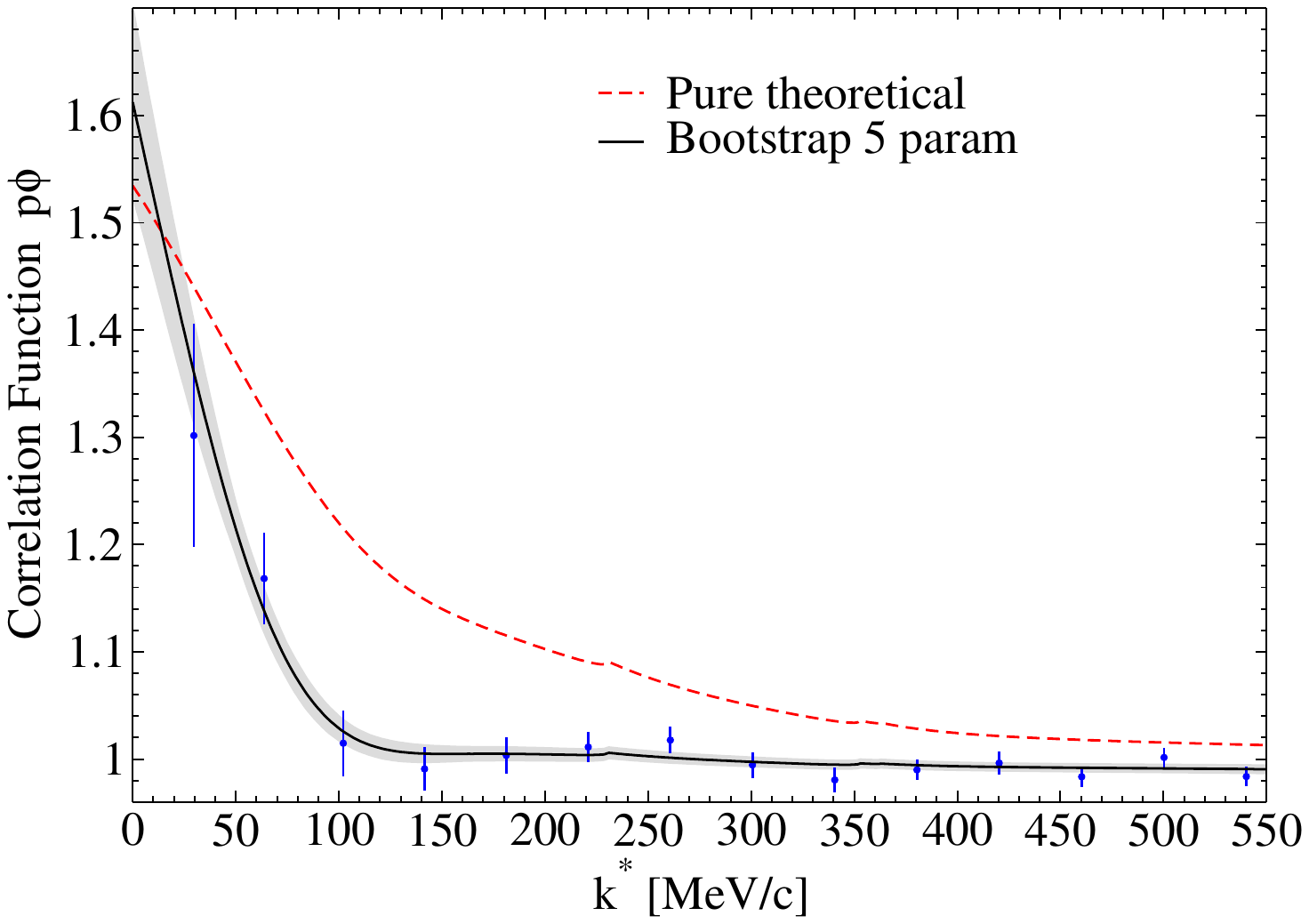}
\vspace{-0.6cm}
\caption{\label{fig:phip_CF} $\phi$p CF for the Pure theoretical model (dashed line) and for the Bootstrap model (solid line), as well as the error band associated (gray shaded band). The experimental points are taken from \cite{ALICE:2021cpv}.}
\end{figure}
\begin{figure}[!ht]
\centering
\includegraphics[width=3.8 in]{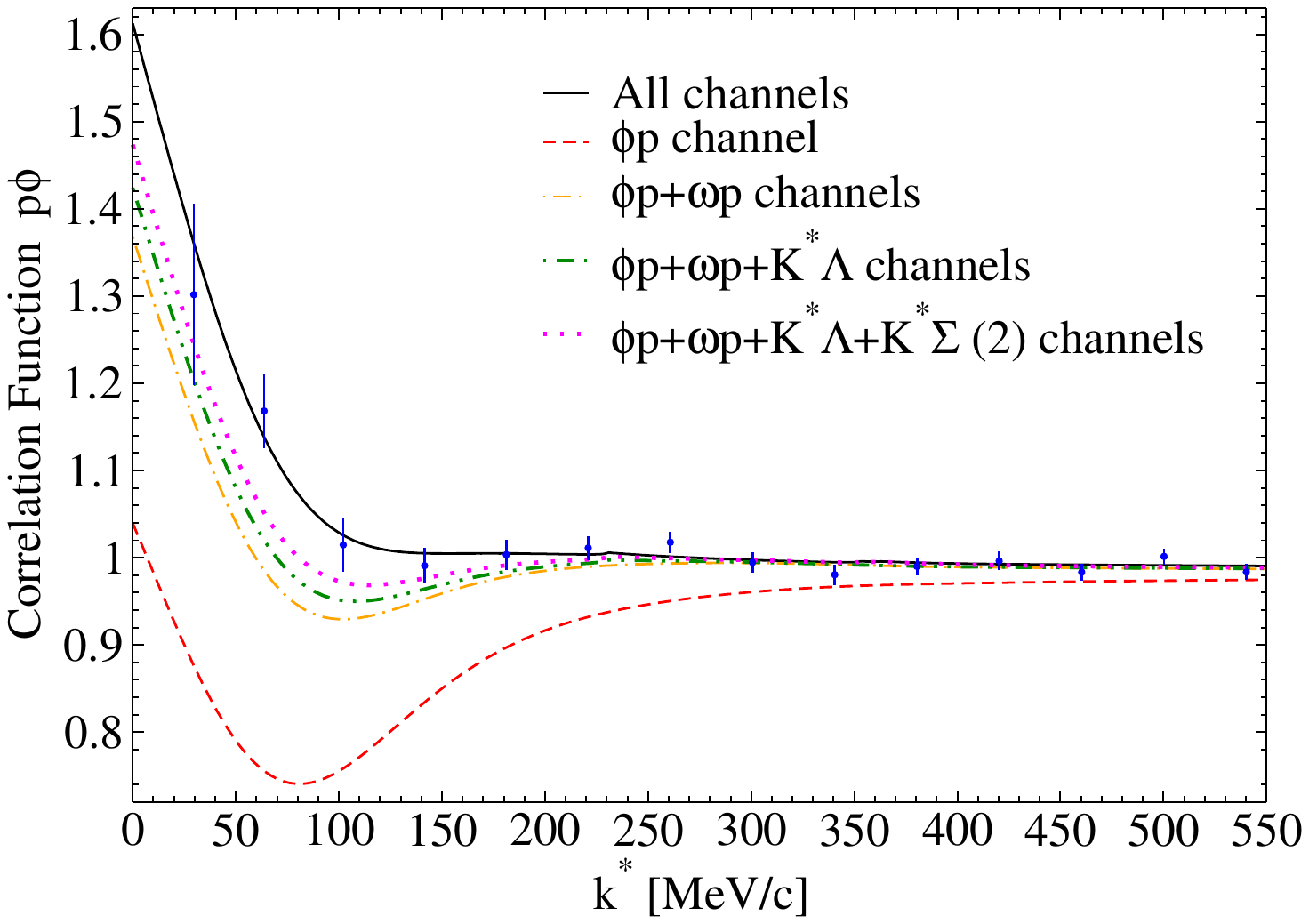}
\vspace{-0.6cm}
\caption{\label{fig:CF_phip_coupled_chan} Contribution of the different transitions to the $\phi$p CF.}
\end{figure}
\begin{table} [h]
  \caption{Effective range, $r_{eff}$ (fm), and scattering length, $a_0$ (fm), for the $\phi$p and  $\rho^0$p channels.} 
  \label{tab:scatt_param}
\centering
\begin{tabular}{lcc}
\hline \\[-2.5mm]
    & \textbf{Pure theoretical}  &  \textbf{Bootstrap}        \\                              
\hline \\[-2.5mm]

$a^{\phi p}_{0}$     & $0.272 +i\, 0.189 $      & $ (-0.034 \pm 0.035) + i\, (0.57 \pm 0.09) $     \\ [1.5mm]
$r^{\phi p}_{eff}$   & $ -7.20 - i\, 0.09$   & $ (-8.06 \pm 2.57 ) + i\, (0.05 \pm 0.53)  $    \\ [1.5mm]
$a^{\rho^0 p}_{0}$     & $0.090 +i\, 0.568 $      & $ (0.09 \pm 0.03) + i\, (0.56 \pm 0.05) $        \\ [1.5mm]
$r^{\rho^0 p}_{eff}$   & $ -3.01 + i\, 98.39$   & $ (-3.05 \pm 0.28 ) + i\, (98.40 \pm 0.12)  $       \\ [2.5mm]

\hline
\end{tabular}
\end{table}

In the first place, in Fig.~\ref{fig:phip_CF}, we show the $\phi$p CF obtained from the two models. The dashed line representing the Pure theoretical model cannot provide a good description of the data yet reproduces the trend and it is of the same order of the experimental data, which gives us confidence in the soundness of the adopted approach. As expected, the CF corresponding to the bootstrap model reaches a good agreement with the data. An interesting feature coming from the cc effects and seen in the CFs obtained from the models is the signature cusp structure emerging due to the openings corresponding to the $K^{*+}\Lambda$ channel, around $230$ MeV/c, and the $K^{*}\Sigma$ channels approximately at $360$ MeV/c. Although, the experimental data have not yet reached the necessary precision to resolve these openings, the predicted strength for them is consistent with the experimental points, therefore the effects of cc in this CF cannot be ruled out.  
A more conclusive proof of the cc relevance in the CF can be found in Fig.~\ref{fig:CF_phip_coupled_chan}, where we show the role of each transition $\psi_{j,\phi p} (k^*,r^*)$ in the $C_{\phi p}(k^*)$. We start computing $C_{\phi p}(k^*)$ taking into account only the elastic $\psi_{\phi p,\phi p}$ (dashed line) and, following Eq.~(\ref{CF}), we progressively add the other channel contributions. From the current figure, it can be immediately seen that all channels contribute to the total $C_{\phi p}(k^*)$ to a greater or lesser extent depending on their interplay and on the penalizing production weights, which are notably lower for the heavier channels compared to light VB pairs (see right column of Table~\ref{production_weights}). Actually, the elastic transition by itself is far from reproducing the experimental data and, consequently, completely different values for the $\phi$p scattering parameters are expected in contrast to those of \cite{ALICE:2021cpv}. The $\phi$p effective range and scattering length extracted from the models are compiled in Table~\ref{tab:scatt_param}. If one directly inspects the real part of $a^{\phi p}_{0}$ for the constrained model, the value is compatible with $0$ within the quoted uncertainties. This result is consistent with the null elastic $C_{ii}$ coefficient of the interaction kernel (see Eq.~(\ref{Kernel}) and Table~\ref{Coeff_WT}). Despite being in tension with the scattering parameters~\cite{ALICE:2021cpv} (see Eq.~\eqref{scatt_ALICE}), the first interpretation of the data, our result is in much better agreement with previous studies: QCD sum rule analysis with different inputs $a^{\phi p}_{0}=-0.01+i\,0.08$ fm \cite{Klingl:1997kf} and $a^{\phi p}_{0}=-0.15\pm 0.02$ fm \cite{Koike:1996ga}, LEPS measurements $a^{\phi p}_{0}=0.15$ fm \cite{Chang:2007fc}, and CLAS data analysis $|a^{\phi p}_{0}|=0.063\pm 0.010$ fm \cite{Strakovsky:2020uqs}. Regarding its imaginary part, the relatively large value acquired is due to the input from the loop functions of the three open channels ($\rho^0$p, $\rho^+$n, and $\omega$p) below the $\phi$p threshold.\\
\begin{figure}[!ht]
\centering
\includegraphics[width=3.8 in]{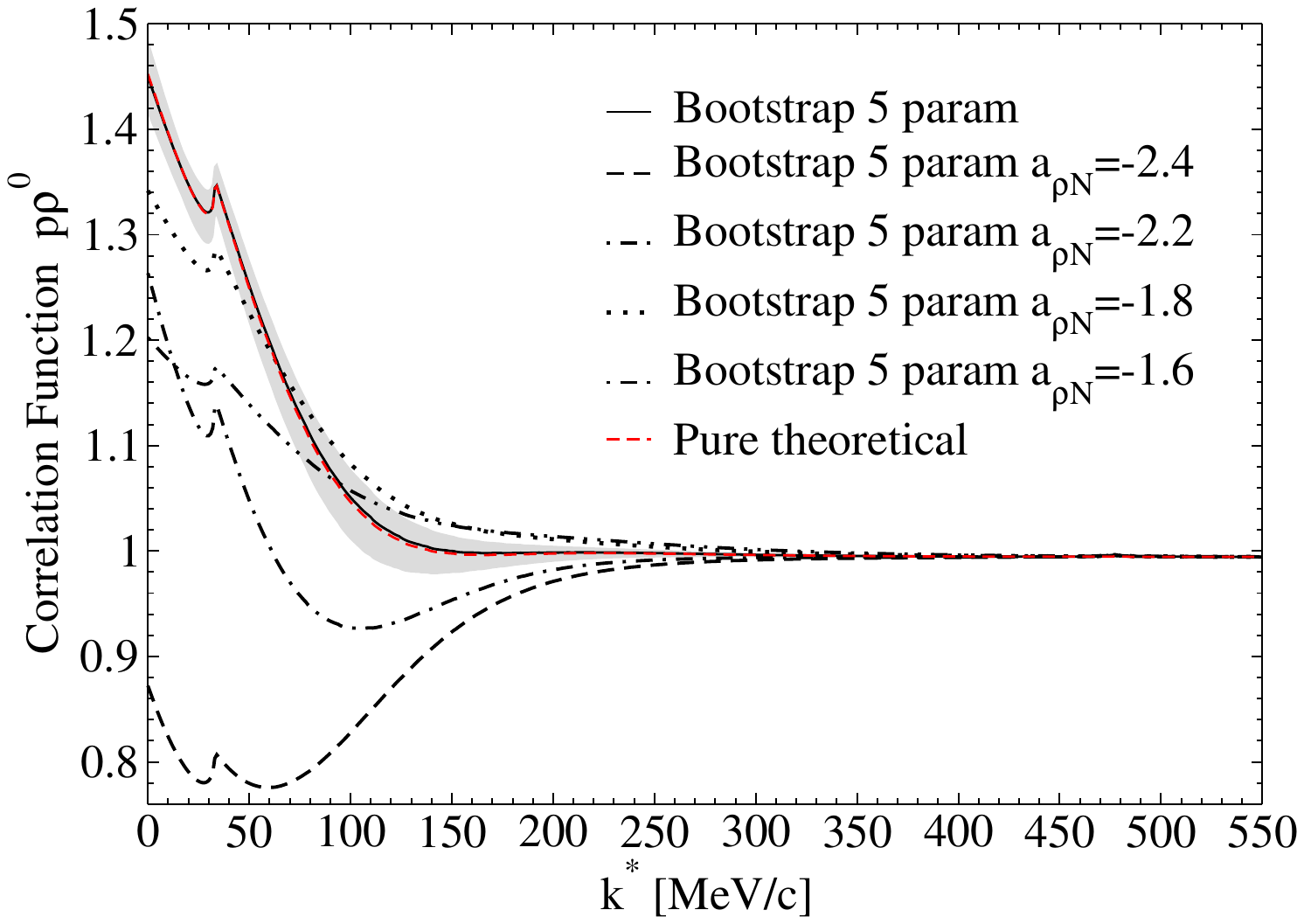}
\vspace{-0.6cm}
\caption{\label{fig:CF_rhop_ND_1} $\rho^0$p CF for the Pure theoretical model (dashed line) and for the Bootstrap model (solid line), as well as the extrapolated error band.}
\end{figure}
\begin{figure}[!ht]
\centering
\includegraphics[width=3.8 in]{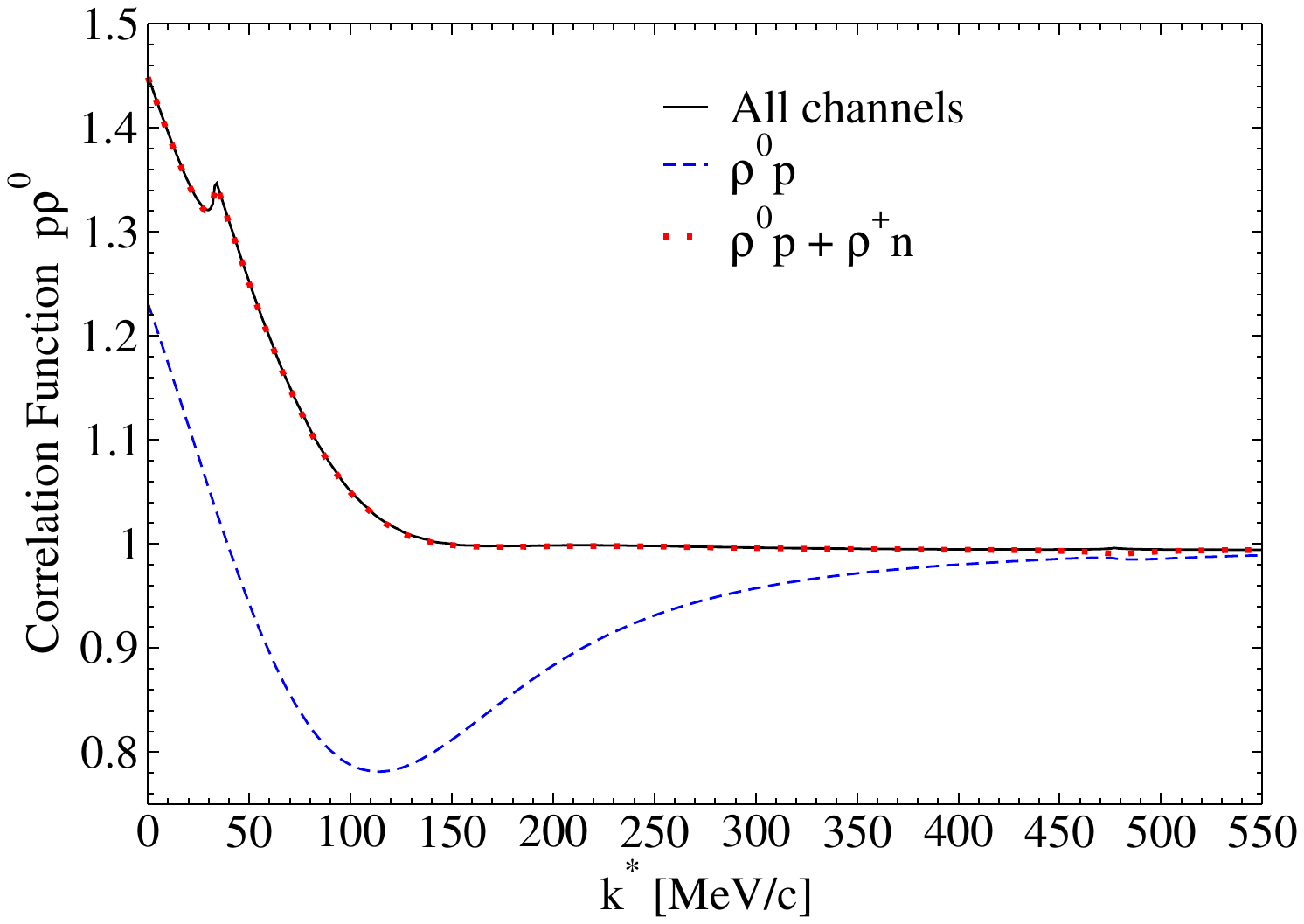}
\vspace{-0.6cm}
\caption{\label{fig:CF_rhop_coupled_chan} Contribution of the different transitions to the $\rho^0$p CF.}
\end{figure}

Motivated by the planned analysis of the $\rho^0$p CF, it seems timely to provide some predictions on this CF. Thus, we use the scattering amplitudes from both models and compute the corresponding $C_{\rho^0 p}(k^*)$. Before continuing with the discussion, it should be commented that, since we have no $\rho^0$p CF data, we cannot have an exact value for the global normalizing factor $N_D$ appearing in Eq.~(\ref{CF_gen}) and, thus, we safely fix it taking $N_D=1$. Our predictions for the $\rho^0$p CF are plotted in Fig.~\ref{fig:CF_rhop_ND_1}, where the most eye-catching feature is that the curve from the Bootstrap model (solid line) perfectly matches the curve for the Pure theoretical model (dashed line). This is a revealing result, because it is demonstrating the dominance of the $N^*(1700)$ state in the $\rho^0$p CF. From \cite{Oset:2010tof}, we already knew that this state is dynamically generated in the $\rho^+$n channel around the $\rho$N thresholds. In the present study, for the reasons already mentioned, we kept $a_{\rho N}$ with the natural size value $-2$ for both models which forces both models to have a coincident position of the pole since the interaction kernel is unchanged. In order to demonstrate it, we decided to vary the $a_{\rho N}$ SC within $10\%$ and $20\%$ in both directions {\footnote {We would like to stress that the $\phi$p CF remains within the error bands of Fig.~\ref{fig:phip_CF} under such variations of $a_{\rho N}$.}}. The effects of these variations are appreciated in Fig.~\ref{fig:CF_rhop_ND_1}, which provide completely different shapes for the $\rho^0$p CF due to the pole location shifts introduced by the new $a_{\rho N}$'s. With these indications, it is clear that the $\rho$N channels have a leading function in the $\rho^0$p CF, but one might wonder if any other channel of the basis can contribute even moderately to it. Bearing this in mind, we proceed as in the previous case, meaning that we start calculating the $C_{\rho^0 p}(k^*)$ just taking into account the elastic transition and we incorporate gradually the other transitions as illustrated in Fig.~\ref{fig:CF_rhop_coupled_chan}. This figure clearly shows the unique dependence of the CF on the $\rho$N transitions. With all previous information, obviously, we cannot give any other prediction for the  $\rho^0$p CF than a theoretical estimate since the $a_{\rho N}$ parameter cannot be constrained by the $\phi$p CF data. Another direct consequence of this fact is the lack of predictive power of the Bootstrap model for $\rho^0$p scattering parameters. From Table~\ref{tab:scatt_param}, it can be seen that both models gives the same values with prominent imaginary part for the scattering length, which is an effect of the convolution of the mass distribution with the loop function to take into account the $\rho$ width. \\
\begin{figure}[!ht]
\centering
\includegraphics[width=3.6 in]{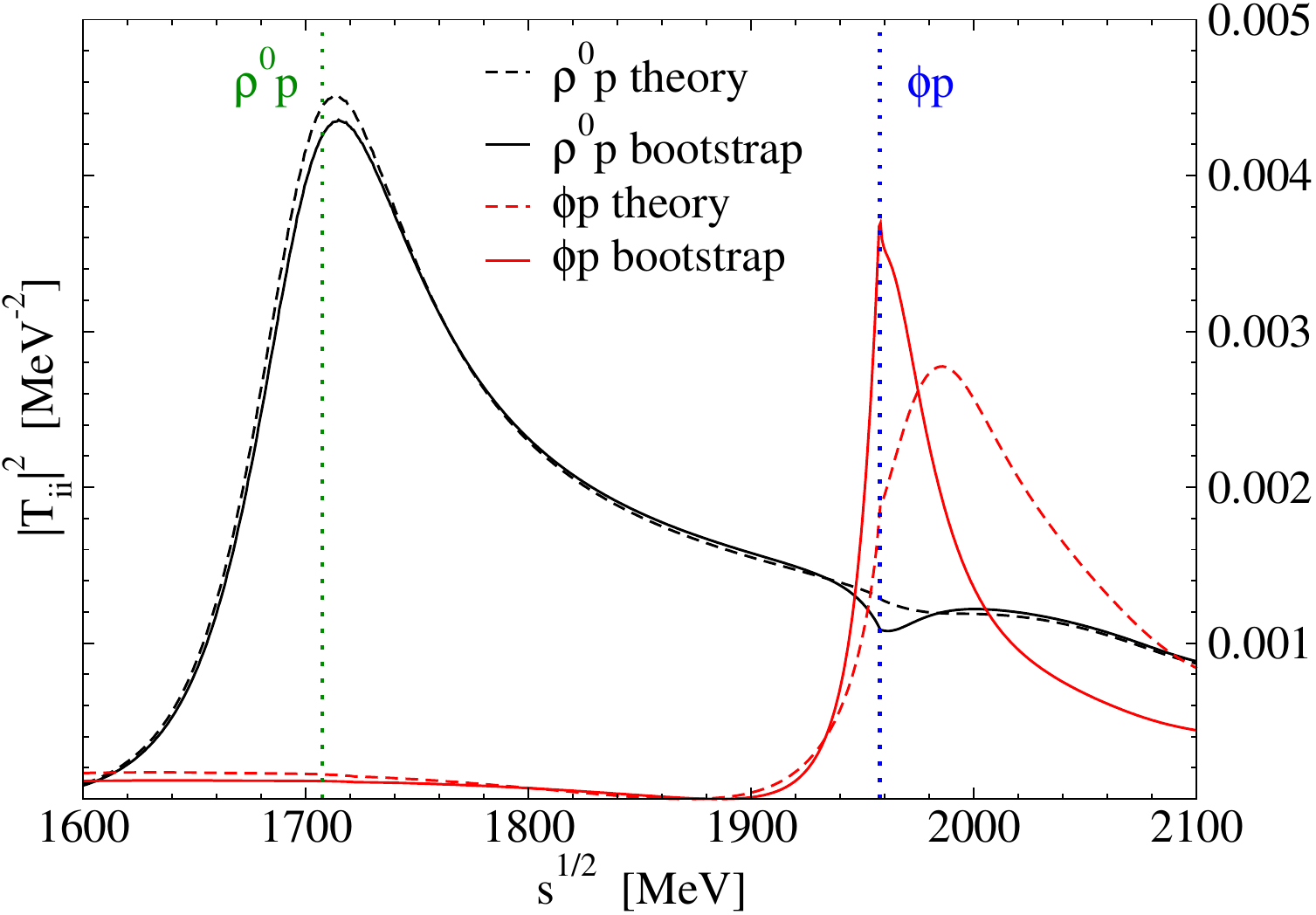}
\vspace{-0.3cm}
\caption{\label{fig:T2} Square of the absolute values for $\rho^0$p and $\phi$p elastic amplitudes obtained in both models: Pure theoretical (dashed lines) and Bootstrap (solid lines). The vertical dotted lines represent the $\rho^0$p and $\phi$p thresholds.}
\end{figure}

\begin{table}[t]
  \caption{Comparison of the pole positions between the models with their couplings  $g_i$ and the corresponding modulus found in $J^P=\frac{1}{2}^-,\frac{3}{2}^-$, $(I,S)=(\frac{1}{2},0)$.}
  \smallskip
  \label{tab:spectroscopy}
\centering
\begin{tabular}{c|cc|cc}
\hline
\hline 
   Model &    \multicolumn{2}{c|}{Pure Theoretical}    &  \multicolumn{2}{c}{Bootstrap} \\
 \hline
  
$M\;\rm[MeV]$           & \multicolumn{2}{c|}{$1977$}      & \multicolumn{2}{c}{$1959$}     \\
$\Gamma/2\;\rm[MeV]$      & \multicolumn{2}{c|}{$52$}       & \multicolumn{2}{c}{$23$}   \\
\hline
                         &   $ g_i$          &   $|g_i|$      &  $g_i$      & $|g_i|$        \\
$\rho$N          & $-0.21-i0.54$     &  $0.58$         &   $-0.08-i0.46$  &  $0.47$   \\
$\omega N$ &  $-1.01-i0.58$    &  $1.17$  &  $-0.68-i0.22$  & $0.72$   \\
$\phi N$ &  $1.39+i0.80$ &  $1.61$  & $0.94+i0.31$  &  $0.99$   \\
$K^{*}\Lambda$  & $2.21-i0.54$  &  $2.28$  &   $1.98-i0.20$  &  $2.00$   \\
$K^{*}\Sigma$  &  $3.75+i0.79$ &  $3.83$  &  $2.95+i0.52$    & $3.00$   \\
\hline 
$M\;\rm[MeV]$    & \multicolumn{2}{c|}{$1700$}      & \multicolumn{2}{c}{$1700$}     \\
$\Gamma/2\;\rm[MeV]$  & \multicolumn{2}{c|}{$-$} & \multicolumn{2}{c}{$-$}   \\
\hline
               &   $ g_i$ & $|g_i|$  &  $g_i$ & $|g_i|$   \\
$\rho$N       & $3.21+i0.00$    &  $3.21$    & $3.22+i0.00$  &  $3.22$   \\
$\omega N$     &  $0.13+i0.00$   &  $0.13$    & $0.11+i0.00$  & $0.11$   \\
$\phi N$       &  $-0.17+i0.00$   &  $0.17$    & $-0.15+i0.00$  & $0.15$   \\
$K^{*}\Lambda$ & $2.32+i0.00$    &  $2.32$    & $2.22+i0.00$  &  $2.22$   \\
$K^{*}\Sigma$  &  $-0.59+i0.00$   &  $0.59$    & $-0.67+i0.00$  & $0.67$   \\

\hline
\hline

\end{tabular}
\end{table}

Finally, we analyze the pole content of the scattering amplitudes. To aid the visualization of the states found with both models, we represent the $|T_{ii}|^2$ for the channels of interest, i. e. $\rho^0$p and $\phi$p, in Fig.~\ref{fig:T2}. Firstly, it can be observed the production of two states each of which located in the neighborhood of the thresholds $\rho^0$p and $\phi$p. If we focus on the state almost located on the $\rho^0$p threshold, we barely appreciate any difference between the $2$ models considered for the reasons already discussed in the former paragraph. This is not the case when one turns the attention to the $\phi$p amplitudes, where it can be noted how the influence of the femtoscopic data has shifted the state around $20$ MeV towards the $\phi$p threshold with respect to the state obtained from the Pure theoretical model. For a deeper understanding of such states, we analytically extrapolate our search for poles in the $T$-matrix complex plane in the proper Riemann sheets. It is convenient to study the scattering amplitudes expressed in isospin basis, this way we can disentangle whether the states are either $N^*\, (I=1/2)$ or $\Delta \, (I=3/2)$ resonances. We did not find any pole in the $I=3/2$ component of the scattering amplitudes, which qualifies the states obtained as $N^*$'s resonances. The resulting pole positions of the resonances with degenerate $J^P= {1/2}^-,{3/2}^-$, together with their couplings $g_i$ to the different channels, are compiled in Table~\ref{tab:spectroscopy} for the Pure theoretical (left panels) and the Bootstrap (right panels) models. The pole lying at higher energies from the Bootstrap model (top right panel) is the one providing new insights. Its location slightly above the $\phi$p threshold seems to naturally reach an accommodation in between all former studies. Being accurate, the analysis of the pole position in the physical basis places it at $1957.75$ MeV, just $20$ KeV above the $\phi$p-channel nominal mass. Obviously, the undergone shift with respect the same state from the Pure theoretical model (in agreement with \cite{Oset:2010tof}) involves a reduction of the width in more than a factor $2$ due to the fact that the accessible phase space with which this state can decay has been decreased. In addition, because of the proximity of the state to the $\phi$p threshold, a Flatté effect \cite{Flatte:1976xu} is also contributing to the diminution of the width, leaving it to an effective value around $37$ MeV, as can be seen in Fig.~\ref{fig:T2}. The pattern of the couplings is similar to those obtained by the previous works using similar approach \cite{Oset:2010tof,Khemchandani:2011et,Gamermann:2011mq,Garzon:2012np}, thus our interpretation of the nature of this state is the same. Regarding the $N^*(1700)$, we cannot provide novel information compared to that of \cite{Oset:2010tof,Garzon:2012np}. Nevertheless, given the constraints provided by the femtoscopic data to the $N^*(1958)$, one can certainly expect to get very valuable information from the ongoing measurement of the $\rho^0$p CF. Here, we would like to comment on the way to get the position of the $N^*(1700)$ pole. Since this state is generated by the $\rho^+$n channel, and also strongly coupled to $\rho^0$p, the width of the $\rho$ vector accounted for in the dressed propagator significantly hampers the pole search and, hence, we proceeded as it is described in Sec. \ref{sec:formalism}. As final remark, in accordance to \cite{Khemchandani:2011et,Gamermann:2011mq}, a second
pole, with a substantially large width and mostly coupling to the $K^*\Sigma$ channels, was found at $2014+i310$ MeV. This large $\Gamma$ causes it to be completely diluted in the real axis. To be more precise, in \cite{Khemchandani:2011et}, the authors managed to reduce drastically the width of this state by incorporating additional contact, s- and u-channel diagrams that favor an extra interplay among channels.

\section{Conclusions}
\label{sec:conclusions}

 We have studied the VB interaction in the ($S=0$, $Q=+1$) sector within a unitary extension of the hidden gauge formalism in coupled channels. The novel aspect of the present study is the use of the $\phi$p CF data to constrain the theoretical model for the first time. The resulting model allowed us to reproduce the data in close agreement with the experimental analysis. From the constrained elastic  amplitude, we extracted new values for the $\phi$p scattering parameters. \\
 
 Another novelty that comes out from the combination of this theoretical approach with femto data is the dynamical generation of a molecular state with $J^P=1/2^-,3/2^-$ slightly above the $\phi$p threshold ($20$ KeV), which is in contrast to previous works within similar approaches that generated it tens of MeV's above. The other pole, historically generated around $1700$ MeV, appeared at the same location since the relevant parameter, i. e. $a_{\rho N}$, to pin such state down turned out not to be constrained by the femtoscopic data. A direct consequence of the previous result is the impossibility of providing a reliable prediction for the $\rho^0$p CF since the $\rho$N amplitudes, the only needed for the case, are dominated by the presence of the $N^*(1700)$ resonance. However, the results for the $\phi$p CF present in this study leads one to think that the ongoing $\rho^0$p CF analysis will certainly help in order obtain a better understanding of the VB dynamics and will provide crucial information about the low-lying pole.\\ 
 
 We also conducted a dedicated study on the relevance of the coupled-channel dynamics in the analysis of the CF data. We demonstrated that the inelastic transitions considered by theory to be relevant in the calculation of both CFs do play an important role, thus providing certainty of the need for these schemes in order to perform a proper analysis of the femtoscopic data.

\section*{Acknowledgements}
\label{sec:Acknowledgements}
The authors are very grateful to J. Nieves and I. Vidaña for the fruitful discussions and their careful reading of the manuscript. This work was supported by ORIGINS cluster DFG under Germany’s Excellence Strategy-EXC2094 - 390783311 and the DFG through the Grant SFB 1258 ``Neutrinos and Dark Matter in Astro and Particle Physics”.

 \bibliographystyle{elsarticle-num} 
 \bibliography{cas-refs}

\begin{thebibliography}{10}
\expandafter\ifx\csname url\endcsname\relax
  \def\url#1{\texttt{#1}}\fi
\expandafter\ifx\csname urlprefix\endcsname\relax\def\urlprefix{URL }\fi
\expandafter\ifx\csname href\endcsname\relax
  \def\href#1#2{#2} \def\path#1{#1}\fi

\bibitem{ALICE:2021cpv}
S.~Acharya, et~al., {Experimental Evidence for an Attractive p-$\phi$
  Interaction}, Phys. Rev. Lett. 127~(17) (2021) 172301.
\newblock \href {http://arxiv.org/abs/2105.05578} {\path{arXiv:2105.05578}},
  \href {https://doi.org/10.1103/PhysRevLett.127.172301}
  {\path{doi:10.1103/PhysRevLett.127.172301}}.

\bibitem{Lednicky:1981su}
R.~Lednicky, V.~L. Lyuboshits, {Final State Interaction Effect on Pairing
  Correlations Between Particles with Small Relative Momenta}, Yad. Fiz. 35
  (1981) 1316--1330.

\bibitem{Chizzali:2022pjd}
E.~Chizzali, Y.~Kamiya, R.~Del~Grande, T.~Doi, L.~Fabbietti, T.~Hatsuda,
  Y.~Lyu, {Indication of a p\textendash{}\ensuremath{\phi} bound state from a
  correlation function analysis}, Phys. Lett. B 848 (2024) 138358.
\newblock \href {http://arxiv.org/abs/2212.12690} {\path{arXiv:2212.12690}},
  \href {https://doi.org/10.1016/j.physletb.2023.138358}
  {\path{doi:10.1016/j.physletb.2023.138358}}.

\bibitem{PhysRevD.106.074507}
Y.~Lyu, T.~Doi, T.~Hatsuda, Y.~Ikeda, J.~Meng, K.~Sasaki, T.~Sugiura,
  \href{https://link.aps.org/doi/10.1103/PhysRevD.106.074507}{Attractive
  $n\text{\ensuremath{-}}\ensuremath{\phi}$ interaction and two-pion tail from
  lattice qcd near physical point}, Phys. Rev. D 106 (2022) 074507.
\newblock \href {https://doi.org/10.1103/PhysRevD.106.074507}
  {\path{doi:10.1103/PhysRevD.106.074507}}.
\newline\urlprefix\url{https://link.aps.org/doi/10.1103/PhysRevD.106.074507}

\bibitem{Huang:2005gw}
F.~Huang, Z.~Y. Zhang, Y.~W. Yu, {N phi state in chiral quark model}, Phys.
  Rev. C 73 (2006) 025207.
\newblock \href {http://arxiv.org/abs/nucl-th/0512079}
  {\path{arXiv:nucl-th/0512079}}, \href
  {https://doi.org/10.1103/PhysRevC.73.025207}
  {\path{doi:10.1103/PhysRevC.73.025207}}.

\bibitem{Belyaev:2007yc}
V.~B. Belyaev, W.~Sandhas, I.~I. Shlyk, {New nuclear three-body clusters
  \textbackslash{}phi{NN}}, Few Body Syst. 44 (2008) 347--349.
\newblock \href {http://arxiv.org/abs/0707.4615} {\path{arXiv:0707.4615}},
  \href {https://doi.org/10.1007/s00601-008-0324-5}
  {\path{doi:10.1007/s00601-008-0324-5}}.

\bibitem{Sofianos_2010}
S.~A. Sofianos, G.~J. Rampho, M.~Braun, R.~M. Adam,
  \href{https://dx.doi.org/10.1088/0954-3899/37/8/085109}{The $\phi$nn and
  $\phi\phi$-nn mesic nuclear systems}, Journal of Physics G: Nuclear and
  Particle Physics 37~(8) (2010) 085109.
\newblock \href {https://doi.org/10.1088/0954-3899/37/8/085109}
  {\path{doi:10.1088/0954-3899/37/8/085109}}.
\newline\urlprefix\url{https://dx.doi.org/10.1088/0954-3899/37/8/085109}

\bibitem{Gao:2017hya}
H.~Gao, H.~Huang, T.~Liu, J.~Ping, F.~Wang, Z.~Zhao, {Search for a hidden
  strange baryon-meson bound state from \ensuremath{\phi} production in a
  nuclear medium}, Phys. Rev. C 95~(5) (2017) 055202.
\newblock \href {http://arxiv.org/abs/1701.03210} {\path{arXiv:1701.03210}},
  \href {https://doi.org/10.1103/PhysRevC.95.055202}
  {\path{doi:10.1103/PhysRevC.95.055202}}.

\bibitem{Sun:2022cxf}
B.-X. Sun, Y.-Y. Fan, Q.-Q. Cao, {The \ensuremath{\phi} p bound state in the
  unitary coupled-channel approximation}, Commun. Theor. Phys. 75~(5) (2023)
  055301.
\newblock \href {http://arxiv.org/abs/2206.02961} {\path{arXiv:2206.02961}},
  \href {https://doi.org/10.1088/1572-9494/acc31d}
  {\path{doi:10.1088/1572-9494/acc31d}}.

\bibitem{Oset:2010tof}
E.~Oset, A.~Ramos, {Dynamically generated resonances from the vector
  octet-baryon octet interaction}, Eur. Phys. J. A 44 (2010) 445--454.
\newblock \href {http://arxiv.org/abs/0905.0973} {\path{arXiv:0905.0973}},
  \href {https://doi.org/10.1140/epja/i2010-10957-3}
  {\path{doi:10.1140/epja/i2010-10957-3}}.

\bibitem{Khemchandani:2011et}
K.~P. Khemchandani, H.~Kaneko, H.~Nagahiro, A.~Hosaka, {Vector meson-Baryon
  dynamics and generation of resonances}, Phys. Rev. D 83 (2011) 114041.
\newblock \href {http://arxiv.org/abs/1104.0307} {\path{arXiv:1104.0307}},
  \href {https://doi.org/10.1103/PhysRevD.83.114041}
  {\path{doi:10.1103/PhysRevD.83.114041}}.

\bibitem{Gamermann:2011mq}
D.~Gamermann, C.~Garcia-Recio, J.~Nieves, L.~L. Salcedo, {Odd Parity Light
  Baryon Resonances}, Phys. Rev. D 84 (2011) 056017.
\newblock \href {http://arxiv.org/abs/1104.2737} {\path{arXiv:1104.2737}},
  \href {https://doi.org/10.1103/PhysRevD.84.056017}
  {\path{doi:10.1103/PhysRevD.84.056017}}.

\bibitem{Garzon:2012np}
E.~J. Garzon, E.~Oset, {Effects of pseudoscalar-baryon channels in the
  dynamically generated vector-baryon resonances}, Eur. Phys. J. A 48 (2012) 5.
\newblock \href {http://arxiv.org/abs/1201.3756} {\path{arXiv:1201.3756}},
  \href {https://doi.org/10.1140/epja/i2012-12005-x}
  {\path{doi:10.1140/epja/i2012-12005-x}}.

\bibitem{Bando:1984ej}
M.~Bando, T.~Kugo, S.~Uehara, K.~Yamawaki, T.~Yanagida, {Is rho Meson a
  Dynamical Gauge Boson of Hidden Local Symmetry?}, Phys. Rev. Lett. 54 (1985)
  1215.
\newblock \href {https://doi.org/10.1103/PhysRevLett.54.1215}
  {\path{doi:10.1103/PhysRevLett.54.1215}}.

\bibitem{Bando:1987br}
M.~Bando, T.~Kugo, K.~Yamawaki, {Nonlinear Realization and Hidden Local
  Symmetries}, Phys. Rept. 164 (1988) 217--314.
\newblock \href {https://doi.org/10.1016/0370-1573(88)90019-1}
  {\path{doi:10.1016/0370-1573(88)90019-1}}.

\bibitem{Meissner:1987ge}
U.~G. Meissner, {Low-Energy Hadron Physics from Effective Chiral Lagrangians
  with Vector Mesons}, Phys. Rept. 161 (1988) 213.
\newblock \href {https://doi.org/10.1016/0370-1573(88)90090-7}
  {\path{doi:10.1016/0370-1573(88)90090-7}}.

\bibitem{Harada:2003jx}
M.~Harada, K.~Yamawaki, {Hidden local symmetry at loop: A New perspective of
  composite gauge boson and chiral phase transition}, Phys. Rept. 381 (2003)
  1--233.
\newblock \href {http://arxiv.org/abs/hep-ph/0302103}
  {\path{arXiv:hep-ph/0302103}}, \href
  {https://doi.org/10.1016/S0370-1573(03)00139-X}
  {\path{doi:10.1016/S0370-1573(03)00139-X}}.

\bibitem{Sarti:2023wlg}
V.~M. Sarti, A.~Feijoo, I.~Vida\~na, A.~Ramos, F.~Giacosa, T.~Hyodo, Y.~Kamiya,
  {Constraining the low-energy S=-2 meson-baryon interaction with two-particle
  correlations} (9 2023).
\newblock \href {http://arxiv.org/abs/2309.08756} {\path{arXiv:2309.08756}}.

\bibitem{Sarkar:2010saz}
S.~Sarkar, B.-X. Sun, E.~Oset, M.~J. Vicente~Vacas, {Dynamically generated
  resonances from the vector octet-baryon decuplet interaction}, Eur. Phys. J.
  A 44 (2010) 431--443.
\newblock \href {http://arxiv.org/abs/0902.3150} {\path{arXiv:0902.3150}},
  \href {https://doi.org/10.1140/epja/i2010-10956-4}
  {\path{doi:10.1140/epja/i2010-10956-4}}.

\bibitem{Oller:2000fj}
J.~A. Oller, U.~G. Meissner, {Chiral dynamics in the presence of bound states:
  Kaon nucleon interactions revisited}, Phys. Lett. B 500 (2001) 263--272.
\newblock \href {http://arxiv.org/abs/hep-ph/0011146}
  {\path{arXiv:hep-ph/0011146}}, \href
  {https://doi.org/10.1016/S0370-2693(01)00078-8}
  {\path{doi:10.1016/S0370-2693(01)00078-8}}.

\bibitem{Lisa:2005dd}
M.~A. Lisa, S.~Pratt, R.~Soltz, U.~Wiedemann, {Femtoscopy in relativistic heavy
  ion collisions}, Ann. Rev. Nucl. Part. Sci. 55 (2005) 357--402.
\newblock \href {http://arxiv.org/abs/nucl-ex/0505014}
  {\path{arXiv:nucl-ex/0505014}}, \href
  {https://doi.org/10.1146/annurev.nucl.55.090704.151533}
  {\path{doi:10.1146/annurev.nucl.55.090704.151533}}.

\bibitem{Haidenbauer:2018jvl}
J.~Haidenbauer, {Coupled-channel effects in hadron\textendash{}hadron
  correlation functions}, Nucl. Phys. A 981 (2019) 1--16.
\newblock \href {http://arxiv.org/abs/1808.05049} {\path{arXiv:1808.05049}},
  \href {https://doi.org/10.1016/j.nuclphysa.2018.10.090}
  {\path{doi:10.1016/j.nuclphysa.2018.10.090}}.

\bibitem{Vidana:2023olz}
I.~Vida\~na, A.~Feijoo, M.~Albaladejo, J.~Nieves, E.~Oset, {Femtoscopic
  correlation function for the Tcc(3875)+ state}, Phys. Lett. B 846 (2023)
  138201.
\newblock \href {http://arxiv.org/abs/2303.06079} {\path{arXiv:2303.06079}},
  \href {https://doi.org/10.1016/j.physletb.2023.138201}
  {\path{doi:10.1016/j.physletb.2023.138201}}.

\bibitem{ALICE:2022yyh}
S.~Acharya, et~al., {Constraining the ${\overline{\textrm{K}}}{\textrm{N}}$
  coupled channel dynamics using femtoscopic correlations at the LHC}, Eur.
  Phys. J. C 83~(4) (2023) 340.
\newblock \href {http://arxiv.org/abs/2205.15176} {\path{arXiv:2205.15176}},
  \href {https://doi.org/10.1140/epjc/s10052-023-11476-0}
  {\path{doi:10.1140/epjc/s10052-023-11476-0}}.

\bibitem{Molina:2023jov}
R.~Molina, C.-W. Xiao, W.-H. Liang, E.~Oset, {Correlation functions for the
  N*(1535) and the inverse problem}, Phys. Rev. D 109~(5) (2024) 054002.
\newblock \href {http://arxiv.org/abs/2310.12593} {\path{arXiv:2310.12593}},
  \href {https://doi.org/10.1103/PhysRevD.109.054002}
  {\path{doi:10.1103/PhysRevD.109.054002}}.

\bibitem{Efron:1986hys}
B.~Efron, R.~Tibshirani, {An introduction to the bootstrap}, Statist. Sci.
  57~(1) (1986) 54--75.

\bibitem{Klingl:1997kf}
F.~Klingl, N.~Kaiser, W.~Weise, {Current correlation functions, QCD sum rules
  and vector mesons in baryonic matter}, Nucl. Phys. A 624 (1997) 527--563.
\newblock \href {http://arxiv.org/abs/hep-ph/9704398}
  {\path{arXiv:hep-ph/9704398}}, \href
  {https://doi.org/10.1016/S0375-9474(97)88960-9}
  {\path{doi:10.1016/S0375-9474(97)88960-9}}.

\bibitem{Koike:1996ga}
Y.~Koike, A.~Hayashigaki, {QCD sum rules for rho, omega, phi meson - nucleon
  scattering lengths and the mass shifts in nuclear medium}, Prog. Theor. Phys.
  98 (1997) 631--652.
\newblock \href {http://arxiv.org/abs/nucl-th/9609001}
  {\path{arXiv:nucl-th/9609001}}, \href {https://doi.org/10.1143/PTP.98.631}
  {\path{doi:10.1143/PTP.98.631}}.

\bibitem{Chang:2007fc}
W.~C. Chang, et~al., {Forward coherent phi-meson photoproduction from deuterons
  near threshold}, Phys. Lett. B 658 (2008) 209--215.
\newblock \href {http://arxiv.org/abs/nucl-ex/0703034}
  {\path{arXiv:nucl-ex/0703034}}, \href
  {https://doi.org/10.1016/j.physletb.2007.11.009}
  {\path{doi:10.1016/j.physletb.2007.11.009}}.

\bibitem{Strakovsky:2020uqs}
I.~I. Strakovsky, L.~Pentchev, A.~Titov, {Comparative analysis of $\omega p$,
  $\phi p$, and $J/\psi p$ scattering lengths from A2, CLAS, and GlueX
  threshold measurements}, Phys. Rev. C 101~(4) (2020) 045201.
\newblock \href {http://arxiv.org/abs/2001.08851} {\path{arXiv:2001.08851}},
  \href {https://doi.org/10.1103/PhysRevC.101.045201}
  {\path{doi:10.1103/PhysRevC.101.045201}}.

\bibitem{Flatte:1976xu}
S.~M. Flatte, {Coupled - Channel Analysis of the pi eta and K anti-K Systems
  Near K anti-K Threshold}, Phys. Lett. B 63 (1976) 224--227.
\newblock \href {https://doi.org/10.1016/0370-2693(76)90654-7}
  {\path{doi:10.1016/0370-2693(76)90654-7}}.

\end{thebibliography}





\end{document}